%% file: main.tex
\documentclass[10pt, conference, letterpaper]{IEEEtran}
\usepackage{cite}
\usepackage{amsmath,amssymb,amsfonts}
\usepackage{algorithmic}
\usepackage{graphicx}
\usepackage{textcomp}
\usepackage[dvipsnames]{xcolor}
\def\BibTeX{{\rm B\kern-.05em{\sc i\kern-.025em b}\kern-.08em
    T\kern-.1667em\lower.7ex\hbox{E}\kern-.125emX}}

\usepackage[nohyperlinks,nolist]{acronym}
\usepackage{url}
\usepackage{todonotes}
\usepackage{subcaption}
\usepackage{comment}
\usepackage[per-mode=symbol]{siunitx}
\usepackage{enumitem}
\usepackage{nohyperref}

\usepackage{booktabs}
\usepackage{tabularx}
\newcolumntype{Y}{>{\centering\arraybackslash}X}
\newcolumntype{L}{>{\arraybackslash}X}
\newcolumntype{R}{>{\raggedleft\arraybackslash}X}
\newcolumntype{C}[1]{>{\centering\arraybackslash}p{#1}}
\usepackage{multirow}
\usepackage{textcmds}
\usepackage[absolute,showboxes]{textpos}

\newcommand{\enum}[1]{({\em #1})~}
\newcommand{\one}{\enum{1}}
\newcommand{\two}{\enum{2}}
\newcommand{\three}{\enum{3}}

\newcommand{\stream}[1]{\textcolor{#1}{\textbf{#1}}}

\begin{document}

\bstctlcite{my:BSTcontrol}

\title{Asynchronous Traffic Shaping and Redundancy: Avoiding Unbounded Latencies in In-Car Networks}

\DeclareRobustCommand*{\fixautherrefmark}[1]{%
  \raisebox{0pt}[0pt][0pt]{\textnormal{{\textsuperscript{\tiny ~#1}}}}%
}
\author{\IEEEauthorblockN{Teresa L\"ubeck\fixautherrefmark{1}, Philipp Meyer\fixautherrefmark{1}, Timo H\"ackel\fixautherrefmark{1, 2}, Franz Korf\fixautherrefmark{1}, and Thomas C. Schmidt\fixautherrefmark{1}}\IEEEauthorblockA{\fixautherrefmark{1}~\href{http://www.haw-hamburg.de/ti-i}{\textit{Department of Computer Science}},
\href{http://www.haw-hamburg.de/ti-i}{\textit{Hamburg University of Applied Sciences}}, Germany \\
{\fixautherrefmark{2}~{\textit{Faculty of Computer Science, Dresden University of Technology}, Germany}}\\
\{\href{mailto:teresa.luebeck@haw-hamburg.de}{teresa.luebeck}, \href{mailto:philipp.meyer@haw-hamburg.de}{philipp.meyer}, \href{mailto:timo.haeckel@haw-hamburg.de}{timo.haeckel}, \href{mailto:franz.korf@haw-hamburg.de}{franz.korf}, \href{mailto:t.schmidt@haw-hamburg.de}{t.schmidt}\}@haw-hamburg.de}
}

\maketitle

\setlength{\TPHorizModule}{\paperwidth}
\setlength{\TPVertModule}{\paperheight}
\TPMargin{5pt}
\begin{textblock}{0.8}(0.1,0.02)
     \noindent
     \footnotesize
     If you cite this paper, please use the original reference:
     Teresa L\"ubeck, Philipp Meyer, Timo H\"ackel, Franz Korf, and Thomas C. Schmidt. "Asynchronous Traffic Shaping and Redundancy: Avoiding Unbounded Latencies in In-Car Networks," In: \emph{Proceedings of the 16th IEEE Vehicular Networking Conference (VNC)}. IEEE Press, June 2025.
\end{textblock}

\begin{abstract}
Time-Sensitive Networking enhances Ethernet-based  In-Vehicle Networks (IVNs) with real-time capabilities.
Different traffic shaping algorithms have been proposed for time-critical communication, of which the Asynchronous Traffic Shaper (ATS) is an upcoming candidate.
However, recent research has shown that ATS can introduce unbounded latencies when shaping traffic from non-FIFO systems.
This impacts the applicability of ATS in IVNs, as these networks often use redundancy mechanisms, i.e. Frame Replication and Elimination
for Reliability (FRER), that can cause non-FIFO behavior.
In this paper, we approach  the problem of accumulated delays from ATS by analyzing the scenarios that generate latency and by devising placement and configuration methods for ATS schedulers to prevent this behavior.
We evaluate our approach in a simulation environment and show how it prevents conditions of unbounded delays.
In an IVN simulation case study, we demonstrate the occurrence of unbounded latencies in a realistic scenario and validate the effectiveness of our solutions in avoiding them.
\end{abstract}
\acresetall


\input{acronyms}

\input{01_introduction.tex}

\input{02_background.tex}

\input{03_problemdescription.tex}

\input{04_solutions_v2.tex}

\input{05_conclusion.tex}

%%%%	Bibliography    %%%%
\bibliographystyle{IEEEtran}
\bibliography{local}

\end{document}

%% file: acronyms.tex
\begin{acronym}

    \acro{ADAS}[ADAS]{Advanced Driver Assistance System}
    \acro{ATS}[ATS]{Asynchronous Traffic Shaper}

    \acro{CAN}[CAN]{Controller Area Network}
    \acro{CBS}[CBS]{Credit Based Shaper}
    \acro{cbs}[\textit{cbs}]{CommittedBurstSize}
    \acro{cir}[\textit{cir}]{CommittedInformationRate}
    \acro{E2E}[E2E]{End-to-End}

    \acro{FRER}[FRER]{Frame Replication and Elimination for Reliability}

    \acro{IFG}[IFG]{Inter-Frame Gap}
    \acro{IR}[IR]{Interleaved Regulator}
    \acro{IVN}[IVN]{In-Vehicle Network}

    \acro{mrt}[\textit{mrt}]{MaximumResidenceTime}

    \acro{NC}[NC]{Network Calculus}

    \acro{PEF}[PEF]{Packet Elimination Function}
    \acro{PFR}[PFR]{Per-Flow Regulator}

    \acro{QoS}[QoS]{Quality of Service}

    \acro{TAS}[TAS]{Time Aware Shaper}
    \acro{TDMA}[TDMA]{Time Division Multiple Access}
    \acro{TSN}[TSN]{Time Sensitive Networking}

    \acro{UBS}[UBS]{Urgency Based Shaper}
\end{acronym}

%% file: 01_introduction.tex
\section{Introduction} \label{sec:introduction}

Future \acp{IVN} will rely on a growing number of sensor data streams to support both control and infotainment functions.
These networks must ensure high reliability, provide sufficient bandwidth, and meet strict \ac{E2E} latency requirements. 
Consequently, \acp{IVN} are transitioning from heterogeneous bus topologies to flat Ethernet backbones~\cite{wtm-avnjr-21,psjtx-svtjr-23}. 
Among the emerging technologies, \ac{TSN} is a promising solution to meet the necessary real-time demands of future \acp{IVN}~\cite{hmks-snsti-19,wtm-avnjr-21,psjtx-svtjr-23,mhlks-fsaad-24}.

The IEEE \mbox{802.1Q}~\cite{ieee8021q-22} collection of standards defines \ac{TSN} ingress and egress control mechanisms for traffic shaping, enabling guaranteed \ac{QoS}.
Within \ac{TSN}, the \ac{ATS} offers per-stream traffic shaping based on a token bucket algorithm that performs well for sporadic traffic~\cite{ntaws-pcijr-19}, achieving lower \ac{E2E} latencies than the widely used \ac{CBS}~\cite{flgx-sacjr-20, zlbpy-sttjr-21}, making it a promising candidate for traffic shaping in \acp{IVN}.
Furthermore, \ac{CBS} is vulnerable to burst cascades~\cite{mvghg-vcbjr-23}, which can be prevented by per-stream and per-hop shaping with \ac{ATS}. 

\ac{TSN} modules are building blocks designed to interoperate; modules with similar purposes should, in principle, be interchangeable.
For instance, different traffic shaping algorithms could theoretically be swapped without issue.
However, this is not always the case: Thomas et al. demonstrate that combining \ac{ATS} with the redundancy mechanism \ac{FRER} can cause unbounded latencies~\cite{tml-wcdjr-22}.
Later, they prove that this issue extends to all non-FIFO networks~\cite{tl-ncsjr-24}, including star topologies.
This limitation is particularly critical for future \ac{IVN} topologies, which often adopt a ring backbone to leverage \ac{FRER} for redundancy~\cite{mhlks-fsaad-24}.
The unbounded latencies caused by the interaction between \ac{ATS} and \ac{FRER} render \ac{ATS} unsuitable for such networks.

To address this challenge, a modification to the \ac{ATS} standard has been proposed~\cite{tl-ncsjr-24}. 
While there are currently no commercially available switches supporting \ac{ATS}, a change of the standard would take time and hinder ongoing development posing significant barriers to adoption. 
The combination of \ac{ATS} and \ac{FRER} needs an in-depth analysis, to reveal if critical cases can occur in specific \ac{IVN} setups and how specific \ac{ATS} configurations can prevent unbounded latencies.

In this paper, we propose \ac{ATS} configurations techniques for non-FIFO networks, enabling bounded latencies in scenarios prone to unbounded delays. 
We reproduce the problem identified in~\cite{tl-ncsjr-24} using simulation scenarios and demonstrate that careful placement of \ac{ATS} schedulers and adjustment of parameters can restore bounded latencies. 
Furthermore, we apply our configuration techniques in a realistic \ac{IVN} setup from~\cite{mhlks-fsaad-24}, where the introduction of \ac{ATS} leads to unbounded latencies and our solutions prevent them.

The remainder of this paper is structured as follows:
Section~\ref{sec:background} introduces \ac{ATS}, \ac{FRER} and related work.
Section~\ref{sec:combine} outlines the problem arising from the combination of \ac{ATS} and \ac{FRER}.
Section~\ref{sec:solutions} presents our \ac{ATS} configuration strategies, while Section~\ref{sec:simulation} evaluates their impact through simulations.
Section \ref{sec:car} examines the proposed configurations in a realistic \ac{IVN} scenario.
Finally, Section \ref{sec:conclusion} concludes the paper with an outlook on future work.

%% file: 02_background.tex
\section{Background and Related Work} \label{sec:background}
Growing complexity and bandwidth demands of modern vehicles drive a transition from \aclp{CAN} toward Ethernet-based networks~\cite{wtm-avnjr-21,hmks-stsdn-22,psjtx-svtjr-23,mhlks-fsaad-24}.
In these networks, \ac{TSN}~\cite{ieee8021q-22} enables deterministic latencies by integrating traffic shaping that handles the coexistence of different traffic classes.

\Ac{TSN} has been explored in various automotive contexts, including its integration with software-defined networking~\cite{hmks-stsdn-22} and its role in anomaly detection~\cite{mhrks-nadct-21}. 
The draft \ac{TSN} automotive profile (IEEE 802.1DG~\cite{ieee8021dg-21}) outlines how different modules can be applied in \acp{IVN}, including the combination of traffic shaping mechanisms.
Most existing work on \ac{TSN} in cars focuses on \ac{CBS} and \ac{TAS}~\cite{wtm-avnjr-21,hmks-stsdn-22,psjtx-svtjr-23}, while \ac{ATS} -- as a relatively new addition -- remains largely unexplored in this domain. Similarly, research on \ac{FRER} has primarily addressed industrial applications rather than automotive use cases.

\subsection{Asynchronous Traffic Shaper (ATS)}
\ac{ATS} was first introduced as \ac{UBS} in 2016 \cite{ss-ubsjr-16} and is now part of \ac{TSN}~\cite{ieee8021q-22}.
\ac{ATS} is a per-stream shaping mechanism, based on a token bucket.
In contrast to \ac{CBS}, scheduling takes place at ingress. 
The \ac{ATS} algorithm calculates an \textit{eligibility time} for every incoming frame --- the earliest time it can be transmitted --- using the amount of tokens in the bucket.
The egress queue for \ac{ATS} traffic is ordered by eligibility time.
Frames with a time earlier than or equal to the current time are eligible for transmission.

One or multiple streams can be assigned to an \ac{ATS} scheduler.
\ac{cir} defines the recovery rate of the token count, and \ac{cbs} the maximum amount of tokens in the bucket; the number of tokens on initialization of the algorithm is \ac{cbs}. 

\ac{ATS} schedulers are organized in \textit{scheduler groups}, each group serving streams that arrive on the same port with the same priority.
Thus, in contrast to other \ac{TSN} shapers, \ac{ATS} depends on ingress information.
Each group has a shared \ac{mrt} parameter that defines an upper limit for the eligibility time delta.
When the scheduling of a frame violates the \ac{mrt}, the frame is dropped.
A shared state variable \textit{group eligibility time}, ensures that the order of frames within the same group is preserved by the scheduling.. 

\begin{figure}
    \centering
    \includegraphics[width=1\linewidth, trim=0.6cm 0.7cm 0.6cm 0.7cm, clip=true]{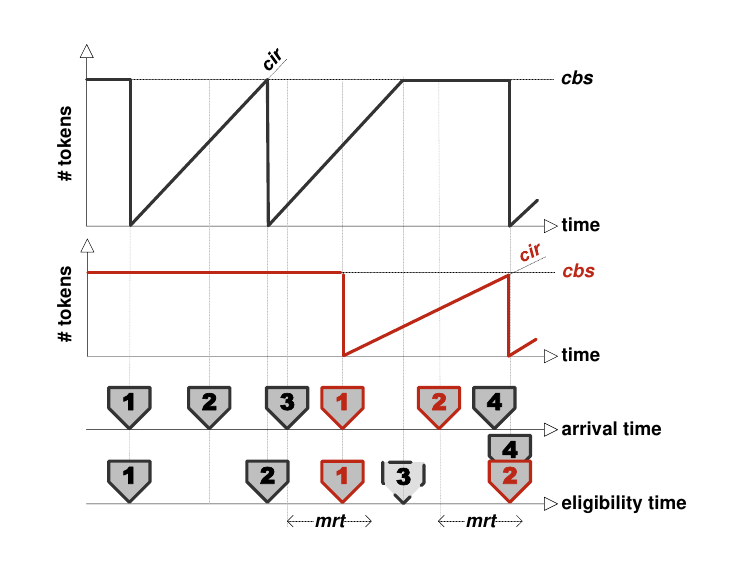}
    \caption{Assignment of frame eligibility times for two streams (\stream{black}, \textcolor{BrickRed}{\textbf{red}}) scheduled by \ac{ATS} in the same scheduler group.}
    \label{fig:ats-numtokens}
\end{figure}

Figure \ref{fig:ats-numtokens} illustrates the assignment of eligibility times for two streams in the same scheduler group, providing an example of the \ac{ATS} algorithm.
Each stream has an associated \ac{ATS} scheduler, represented by the number of tokens in the black and red buckets.
Note that the \ac{cir} and \ac{cbs} values are different between the two schedulers.

First, a frame of the black stream arrives when there are enough tokens in the bucket.
It is assigned its arrival time as eligibility time, and the required tokens are subtracted.
Tokens then recover with rate \ac{cir}. 
The second frame of the black stream arrives when there are not enough tokens available, so its eligibility time is set to the next time when enough tokens accumulate.
The required tokens are subtracted at the time of eligibility.
The assigned eligibility time for the third frame is further in the future than its arrival time + \ac{mrt}, so the frame is dropped and no tokens are subtracted. 

When the first frame of the red stream arrives, there are enough tokens in the bucket to serve it immediately, while the second frame of the red stream must wait.
Scheduler groups are important when the fourth frame of the black stream arrives.
On arrival of the fourth frame, there are enough tokens in the black bucket to serve it immediately, but the group eligibility time is the most recent eligibility time value set by a scheduler in the group.
In this case, the eligibility time of the second frame of the red stream, which is in the future.
The eligibility time of the fourth frame of the black stream is therefore set to the group eligibility time.
The second frame of the red stream and the fourth frame of the black stream have the same eligibility times, but the frame of the red stream arrived first and will be transmitted first. 

\subsection{Frame Replication and Elimination (FRER)}
\ac{FRER} (IEEE 802.1CB~\cite{ieee8021cb-17}) is a redundancy mechanism that enhances reliability by duplicating frames and transmitting them over multiple paths. 
This increases the likelihood that at least one copy reaches the destination. 
Streams can be replicated multiple times within a network and later converge at a merge point, where duplicates are eliminated.

A tag is added to each frame that assigns a sequence number.
A stream splitting function replicates the frames, keeping their sequence number. 
Duplicated frames are transmitted over different paths in the network, ensuring delivery even in the event of a path failure. 
At the receiving end, a recovery function eliminates redundant copies by discarding any duplicate frames with the same sequence number, ensuring that only one instance reaches the destination.

\ac{FRER} does not guarantee that the order of frames in a stream is preserved.
Duplicate frames that take different paths can have different delays and failures on one path can lead to sporadic frame losses.
As an example:
The frame sequence $1, 2, 3$ is split and duplicates $1_l, 2_l, 3_l$ are transmitted over a path with low delay, while duplicates $1_h, 2_h, 3_h$ are transmitted over a path with high delay.
Frames can arrive at the recovery function in the order $2_l, 1_h, 2_h, 3_l, 3_h$ when frame $1_l$ is lost, the frames are then reordered to $2, 1, 3$ after the stream is merged.
Furthermore, frames with different sequence numbers from different paths (e.g., a short and a long path) may arrive at the recovery function simultaneously. 
If none are dropped as duplicates, multiple frames are transmitted together in a burst. 
Repeated occurrences of such bursts can momentarily increase the transmission rate of the stream~\cite[Annex C.9]{ieee8021cb-17}.

\subsection{Related Work}

Several studies compare \ac{ATS} performance with other traffic shaping mechanisms, such as \ac{CBS}, \ac{TAS}, or strict priority shaping~\cite{ntaws-pcijr-19, flgx-sacjr-20, zlbpy-sttjr-21, yi-qeajr-24}, demonstrating its effectiveness in both industrial and \ac{IVN} scenarios.
Some works evaluate combinations of traffic shapers~\cite{msmb-lbbjr-18, ntaws-pcijr-19}.
In contrast, our work does not focus on the relative performance of \ac{ATS} but rather on the feasibility of specific configurations. 
Our goal is to enable the use of \ac{ATS} in realistic \acp{IVN} with redundancy, ensuring its applicability as a viable alternative to other traffic shapers.

The impact of \ac{ATS} parameter configuration is examined in multiple studies:  
Fang et al.~\cite{flgx-sacjr-20} show that \ac{ATS} performance depends on the choice of \ac{cir} and \ac{cbs}. 
If these values are set too low, \ac{ATS} performs worse than \ac{CBS} and strict priority scheduling, and increasing the parameter values improves the performance. 
Hu et al.~\cite{hlxf-dbajr-20} analyze the effect of \ac{cbs} on \ac{E2E} delay bounds, while Yoshimura et al.~\cite{yi-qeajr-24} demonstrate that reducing the \ac{mrt} parameter lowers latency and jitter but increases frame loss rates.
In comparison, we derive a set of configuration guidelines to set \ac{ATS} parameters and evaluate on a binary criterion whether the setting introduces critical delays.
Additionally, we compare the influences of \ac{cir} and \ac{cbs} on the shaping of bursts.

In contrast to other shaping mechanisms, like \ac{CBS}, where the percentage of allocated bandwidth is capped and \textit{class measurement intervals} are provided, there is little guidance on the parametrization of \ac{ATS} in the standard~\cite{ieee8021q-22}.
As a consequence, difficulties in setting up \ac{ATS} have been reported~\cite{ntaws-pcijr-19}.
We observe that many studies use \ac{ATS} parameter values that are higher than the rate and burst of the shaped streams, with the \ac{mrt} parameter often not set at all or assigned a value larger than the simulation time.
This suggests that \ac{ATS} parameters are frequently configured \textit{such that it works}, which may explain why~\cite{tl-ncsjr-24} were the first to report unbounded latencies in non-FIFO networks using \ac{ATS}.
We introduce the problem of unbounded latencies in detail in Section~\ref{sec:combine}.

In this work, we derive and apply configurations to restore bounded latencies. 
we examine synthetic networks where \ac{ATS} parameters are predefined and a realistic \ac{IVN} where some \ac{ATS} parameters are found empirically.

%% file: 03_problemdescription.tex
\section{Unbounded Latencies in ATS Networks} \label{sec:combine}

Network calculus results show that \ac{ATS} does not increase the worst-case latencies of a stream when it is placed behind a FIFO system ~\cite{l-ttrjr-18}.
This does not hold for non-FIFO systems~\cite{tl-ncsjr-24}.
The proof for this is based on an adversarial frame sequence, which leads to unbounded latencies if it is shaped with \ac{ATS}.
We design three synthetic networks that reproduce the adversarial frame sequence, where the non-FIFO behavior is introduced by \ac{FRER} and parallel paths.
We use these networks to evaluate our workaround solutions. 

\begin{figure}
    \centering
    \includegraphics[width=1\linewidth, trim=0.6cm 0.8cm 0.6cm 0.7cm, clip=true]{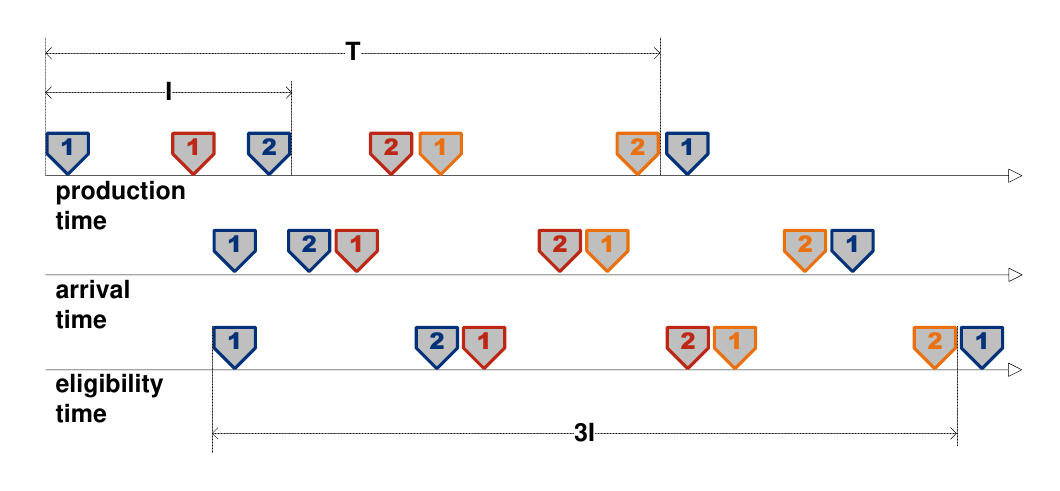}
    \caption{Adversarial frame generation sequence of three concurrent streams (\textcolor{Blue}{\textbf{blue}}, \textcolor{BrickRed}{\textbf{red}}, and \stream{orange}) redrawn from~\cite{tl-ncsjr-24}.}
    \label{fig:ats_unbounded_example}
\end{figure}
\begin{figure}
    \centering
    \subfloat[%Network A: 
    Using \ac{FRER} for redundancy with packet loss.]{\includegraphics[width=1\linewidth, trim=0.75cm 0.6cm 0.7cm 0.6cm, clip=true]{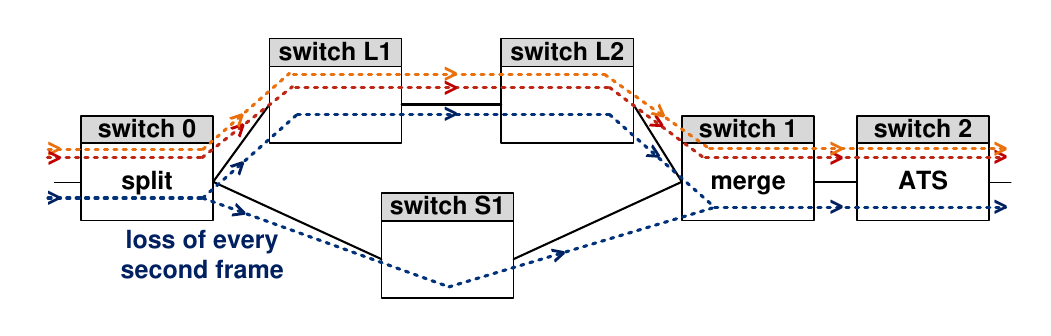}\label{fig:smallnet_frer}}

    \subfloat[%Network B: 
    Using parallel paths in a ring with cross-traffic.]{\includegraphics[width=1\linewidth, trim=0.75cm 0.6cm 0.6cm 0.3cm, clip=true]{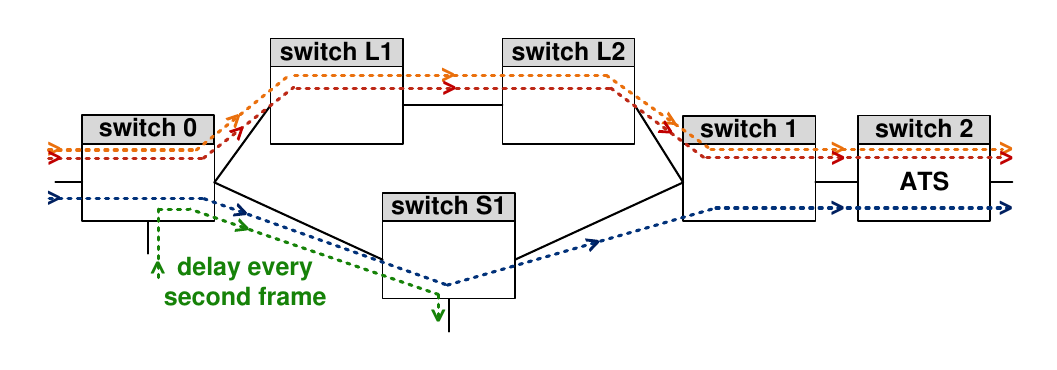}\label{fig:smallnet_circle}}

    \subfloat[%Network C: 
    Using parallel paths in a tree with cross traffic.]{\includegraphics[width=1\linewidth, trim=0.75cm 0.6cm 0.6cm 0.3cm, clip=true]{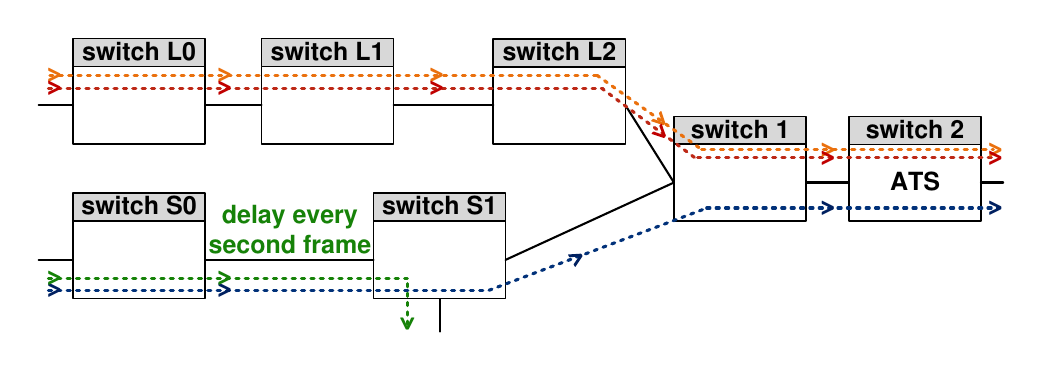}\label{fig:smallnet_parallel}}
    \caption{Switched network with adversarial frame generation (see Fig~\ref{fig:ats_unbounded_example}). End systems are not shown. Each dotted line (\textcolor{Blue}{\textbf{blue}}, \textcolor{BrickRed}{\textbf{red}}, \stream{orange}, and \textcolor{ForestGreen}{\textbf{green}}) represents a different stream.}
    \label{fig:unbounded_networks}
\end{figure}

\begin{figure*}
    \centering
    \includegraphics[width=1\linewidth, trim=0.6cm 0.6cm 0.6cm 0.6cm, clip=true]{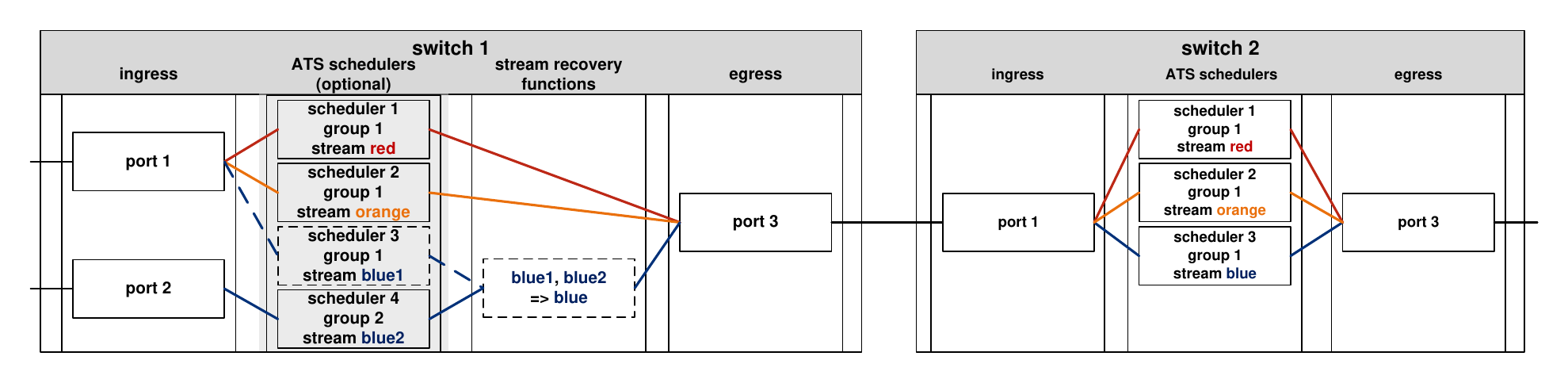}
    \caption{Switch 1 and 2 of the example networks (see Fig.~\ref{fig:unbounded_networks}). Dashed elements appear only in network A, while solid-line elements are present in networks A, B, and C. Colors (\textcolor{Blue}{\textbf{blue}}, \textcolor{BrickRed}{\textbf{red}}, and \stream{orange}) represent the streams.}
    \label{fig:ats_everywhere}
\end{figure*}

Figure \ref{fig:ats_unbounded_example} sketches the problematic frame sequence. 
Three streams, denoted as blue, red, and orange, each produce two frames with a spacing of $I$ in one period.
One period has length $T < 3I$ when the frames are produced at the sources.
Within the network, the second frame of the blue stream overtakes the first frame of the red stream, thus breaking the FIFO property.
At the end, there are three \ac{ATS} schedulers belonging to the same scheduler group, each associated with one stream.
All \ac{ATS} schedulers are configured such that the interval between two outgoing frames of a stream is at least $I$.
Due to the second blue frame arriving before the first red frame the red frame cannot be send before both blue frames are sent, as they are in the same scheduler group. 
This increases the period length after shaping to $3I$.
Repeating the period leads, over time, to infinitely increasing latencies~\cite{tl-ncsjr-24}.

We create three networks where the adversarial frame generation is possible (see Fig.~\ref{fig:unbounded_networks}).
Preconditions for the non-FIFO behavior are fulfilled by two realistic causes:
In network \textbf{A} (see Fig.~\ref{fig:smallnet_frer}) \ac{FRER} with packet loss causes the second blue frame to overtake the first red frame.
In network \textbf{B} (see Fig.~\ref{fig:smallnet_circle}) and \textbf{C} (see Fig.~\ref{fig:smallnet_parallel}) the blue stream is delayed by cross-traffic (green stream).
In all networks, the streams share the same priority and no traffic shapers exist besides \ac{ATS} on switch 2.

The three streams enter network A in their order of production (see Fig.~\ref{fig:ats_unbounded_example}).
The blue stream is split on switch 0 and is transmitted over both a long path (top) and a short path (bottom).
Every second frame on the short path is lost.
The red and orange streams are only transmitted over the long path.
The paths join on switch 1, where the blue stream is merged.
Then all remaining frames enter switch 2 in the arrival sequence (see Fig.~\ref{fig:ats_unbounded_example}), where \ac{ATS} causes latencies to increase.

Network B has the same topology as network A, but without \ac{FRER}.
Instead, the order of frames changes because the blue stream always takes the short path, where every second frame is delayed by a green cross-traffic stream that shares part of the path. 
The red and orange streams take the long path.
When the three streams re-join at switch 1 from different ports, the adversarial frame arrival sequence occurs (see Fig.~\ref{fig:ats_unbounded_example}). 
Again, switch 2 shapes the streams with \ac{ATS}, causing the delays.

Network C shows that the unbounded latencies can occur in networks with a generic star topology.
The blue, red, and orange stream originate from different sources.
The blue stream takes a short path where every second frame is delayed by cross-traffic, while the red and orange streams take a long path.
All three streams meet for the first time on switch 1, where the frames arrive in the adversarial arrival sequence (see Fig.~\ref{fig:ats_unbounded_example}) and are then transmitted to the \ac{ATS} switch.

%% file: 04_solutions_v2.tex
\section{Configuring \ac{ATS} for Bounded Latencies} \label{sec:solutions}
We identify two general approaches to mitigate the described problem of unbounded latencies in non-FIFO networks.
First, the design of the network, i.e., the placement of \ac{ATS} schedulers.
Second, the impact of specific \ac{ATS} configuration parameters.
Sections \textbf{\ref{sec:solution_1}} to \textbf{D} describe our solutions. 

\subsection{Use \ac{ATS} on All Hops}
\label{sec:solution_1}
This approach targets the network design by placing \ac{ATS} schedulers on every switch to break the adversarial arrival sequence.
In the adversarial frame generation, only one switch uses \ac{ATS} (switch 2).
An improved setup re-shapes a stream with \ac{ATS} on every hop.
Specifically, using \ac{ATS} on switch 1 restores the original order of the frames in cases where the non-FIFO property is introduced by parallel paths.
This approach requires additional resources, as all switches need to have traffic shaping configurations.

Figure \ref{fig:ats_everywhere} illustrates the setup of the last two switches in the synthetic networks.
First, the case where the non-FIFO property is due to parallel paths (networks B and C).
The frames of the three streams arrive on switch 1 in the critical order, but the blue stream enters on port 2, while the red and orange streams enter on port 1.
Without \ac{ATS} on switch 1, all frames enter switch 2 in this order, where they are shaped by \ac{ATS} schedulers in the same scheduler group, thus gaining unbounded latencies.
But when \ac{ATS} is also used on switch 1, the streams are shaped there such that their frames arrive on switch 2 in their order of production.
While the second frame of the blue stream is assigned an eligibility time in the future, the first frame of the red stream is assigned its arrival time as eligibility time.
This is due to their schedulers being in different scheduler groups on switch 1.

\begin{figure*}
    \centering

    \subfloat[\ac{cir} and \ac{cbs} set tightly.]{\includegraphics[width=0.3\linewidth, trim=0.6cm 0.7cm 0.6cm 0.6cm, clip=true]{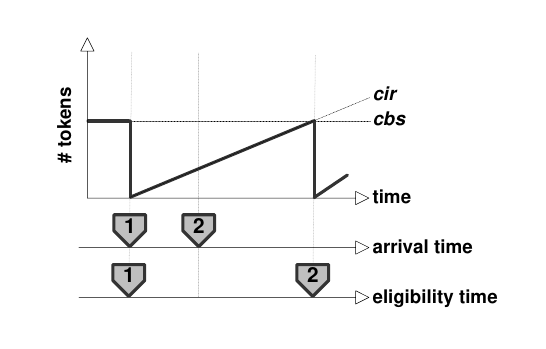}\label{fig:ats_tokens_cir1cbs1}}%
    \subfloat[Value of \ac{cir} is doubled.]{\includegraphics[width=0.3\linewidth, trim=0.6cm 0.7cm 0.6cm 0.6cm, clip=true]{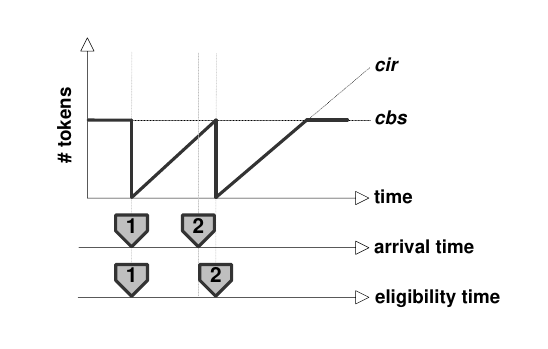}\label{fig:ats_tokens_cir2cbs1}}%
    \subfloat[Value of \ac{cbs} is doubled.]{\includegraphics[width=0.3\linewidth, trim=0.6cm 0.7cm 0.6cm 0.6cm, clip=true]{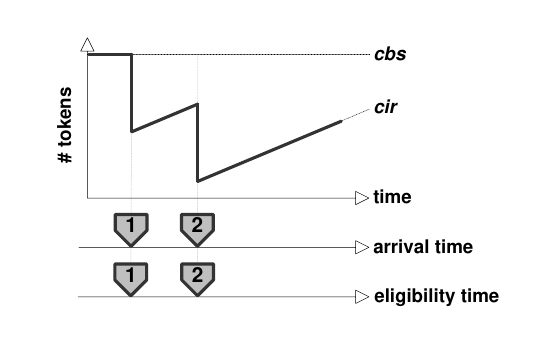}\label{fig:ats_tokens_cir1cbs2}}%
    
    \caption{Qualitative evolution of \ac{ATS} tokens and eligibility times with two consecutively arriving frames of one stream.}
    \label{fig:ats_tokens}
\end{figure*}

This approach does not work when \ac{FRER} is used. 
The reason is the sequence of \ac{ATS} schedulers and \ac{FRER} recovery functions.
Figure \ref{fig:ats_everywhere} shows the setup for network A.
The blue stream enters switch 1 from two different ports and each instance is shaped by its own \ac{ATS} scheduler.
Afterwards, the recovery function drops duplicated frames and causes the critical frame order which is transmitted to switch 2.
Moving the \ac{ATS} schedulers behind the merging does not make sense, due to the definition of scheduler groups.
Scheduler groups require frames to arrive at the same port, but a merged stream has frames arriving on different ports such that a clear assignment to a group is impossible.

\subsection{Increase \ac{cir} or \ac{cbs}}
\label{sec:solution_2}
This following approach is based on increasing the values of \ac{cir} and \ac{cbs} on switch 2 such that the length of the period after shaping is not higher than the length of the period in which frames are produced.
The values of \ac{cir} and \ac{cbs} are set in the adversarial frame generation such that the interval between two frames of a stream is at least $I$ after shaping.
An increase of \ac{cir} or \ac{cbs} reduces the length of this interval, consequently reducing the length of the period after shaping.

The minimum increase of the values can be found by analysis or empirically, and this might not be easy.
But in combination with solution \ref{sec:solution_1}, the increase of the \ac{cir} and \ac{cbs} values must only be done for the \ac{ATS} after the \ac{FRER} merging.
\ac{FRER} can cause a temporary doubled bandwidth \cite[Annex C.9]{ieee8021cb-17}.
So, doubling the values of \ac{cir} and \ac{cbs} on the next \ac{ATS} scheduler after merging, can ensure that the shaping does not add delays.
It is also possible to increase either \ac{cir} or \ac{cbs}.
Doubling only \ac{cir} takes the perspective that the data rate doubles due to \ac{FRER}.
While only increasing \ac{cbs} takes the perspective that \ac{FRER} causes bursts.
To ensure that the burst is not shaped, the value of \ac{cbs} must be set to \textit{burst size + original value of cbs}.

The \ac{ATS} scheduler behaves differently when only \ac{cir} or only \ac{cbs} is increased.
Figure \ref{fig:ats_tokens} shows the number of tokens, arrival and eligibility times for a burst of two frames arriving at an \ac{ATS} scheduler.
Figure \ref{fig:ats_tokens_cir1cbs1} shows the behavior when \ac{cir} is set to the data rate of the stream and \ac{cbs} to one frame size.
The second frame is delayed due to the shaping.
Doubling the value of \ac{cir} alone also results in a delay of the second frame, but the delay is only half as large as before (\ref{fig:ats_tokens_cir2cbs1}).
This is enough to prevent the increase of the period after shaping the adversarial frame generation.
Doubling the value of \ac{cbs} allows both frames to be transmitted without delay (\ref{fig:ats_tokens_cir1cbs2}).

Increasing any of the parameters \ac{cir} or \ac{cbs}, changes the shape of the stream on all following hops. 

\subsection{Do not Place \ac{ATS} Behind \ac{FRER} Merge Points}
\label{sec:solution_3}
For some networks the merging only takes place on the last switch before the destination.
In these cases it can be useful to omit the \ac{ATS} shaping after merging, if necessary replacing it with a different shaping mechanism.
This solution works well in combination with \ac{FRER}, because the merger is a known location where frames can change their order.
When the unbounded latencies are introduced by other means, as in networks B and C, it is more difficult to locate the \ac{ATS} schedulers that can potentially cause unbounded latencies and selectively remove them.
Solution \ref{sec:solution_1} should be used in combination with this solution to prevent unbounded latencies that are not caused by \ac{FRER}.

\subsection{Utilize the \ac{mrt} Parameter}
\label{sec:solution_4}
Setting the \ac{mrt} parameter for the \ac{ATS} scheduler on switch 2, can prevent unbounded latencies by dropping frames.
When a frames eligibility time is set later than its arrival time plus $mrt$, the frame is dropped.
This means for the adversarial frame generation, that the continuous increase of latencies due to shaping is interrupted when the assigned eligibility times of the frames are too far in the future. 
At this point frames are dropped in regular intervals, such that the period length after shaping is not larger than the period length at production.
There is no decrease in latencies, because the arrival of frames stays as is.
This solution only shifts the problem of unbounded latencies towards a problem of frame loss.
\acp{IVN} have a high demand for reliability, making this solution not applicable.

\section{Evaluation: Avoiding  Adversarial Frames} \label{sec:simulation}

To verify that the proposed configurations restore bounded latencies for the adversarial frame generation, we simulate the three networks from Figure \ref{fig:unbounded_networks}.
The simulation is in OMNeT++ 6.0.2 \cite{omnetpp} and uses the INET framework \cite{inet-framework}.
Simulation enables the implementation and evaluation of our proposed solutions without needing to wait for hardware components implementing \ac{ATS} that are not yet commercially available.

\subsection{Baseline Network Setups}

There are three networks that are simulated.
All lines in the networks have a bandwidth of 100\,\unit{\mega\bit/\second}.

Network A (Figure \ref{fig:smallnet_frer}) produces the blue, red, and orange stream on a device connected to switch 0. 
The long path has four switches and the short path one switch.
The listener device is after switch 2.
Network B (Figure \ref{fig:smallnet_circle}) has two talker devices connected to switch 0; one produces the red and orange stream, the other the blue and green stream.
The listener device for the green stream is connected to switch S1.
The length of the paths and the listener device for the blue, red, and orange stream are the same as in network A.
Network C (Figure \ref{fig:smallnet_parallel}) has a long path with five switches and a short path with two switches.
The red and orange stream are produced on a device connected to switch L0, the blue and green stream are produced to a device connected to switch S0.
The listener device for the blue, red and orange stream is after switch 2, while the destination of the green stream is connected to switch S1.

We simulate the adversarial frame generation with the following stream and \ac{ATS} configurations:
The frame size for the blue, red, and orange stream is 125\,\unit{\byte} (1000\,\unit{\bit}) including overhead and \ac{IFG}.
The interval between two frames of a stream that are produced within a period is $I = 50\,\unit{\us}$, a period repeats after $T = 140\,\unit{\us}$.
The offset between the first frame of the blue stream and the first frame of the red stream is $20\,\unit{\us}$, while the offset between the second frame of the red stream and the first frame of the orange stream is $10\,\unit{\us}$.
The cross-traffic frames in networks B and C have a size of $500\,\unit{\byte}$ and are produced every $140\,\unit{\us}$.
\ac{ATS} parameters are the same for all three streams: $\ac{cir} = 20\,\unit{\mega\bit/\second}$ (which is higher than the average bandwidth of the streams), and $\ac{cbs}=125\,\unit{\byte}$ (one frame size), \ac{mrt} is set infinitely high to prevent \ac{ATS} from dropping frames and therefore limiting the delay.
The period length after shaping is $3I = 150\,\unit{\us}$.

We simulate three baseline cases and the four proposed \ac{ATS} configurations for each of the three networks for 10\,\unit{s} each.
Behaviors observed in the simulations do not change with longer simulation times, as the traffic generation and behavior of the modules is static.

\subsection{Evaluation of Baseline Cases}
We first simulate three baseline cases, to show that the problem occurs in our networks due to \ac{ATS} on switch 2.
These are: case \enum{a} without \ac{ATS} on any switch, case \enum{b} with \ac{ATS} only on switch 2, and case \enum{c} with \ac{ATS} only on switch 2, but the schedulers are in different groups.

\begin{figure}
    \centering
    \includegraphics[width=1\linewidth, trim=0.1cm 0.3cm 0.1cm 0.2cm, clip=true]{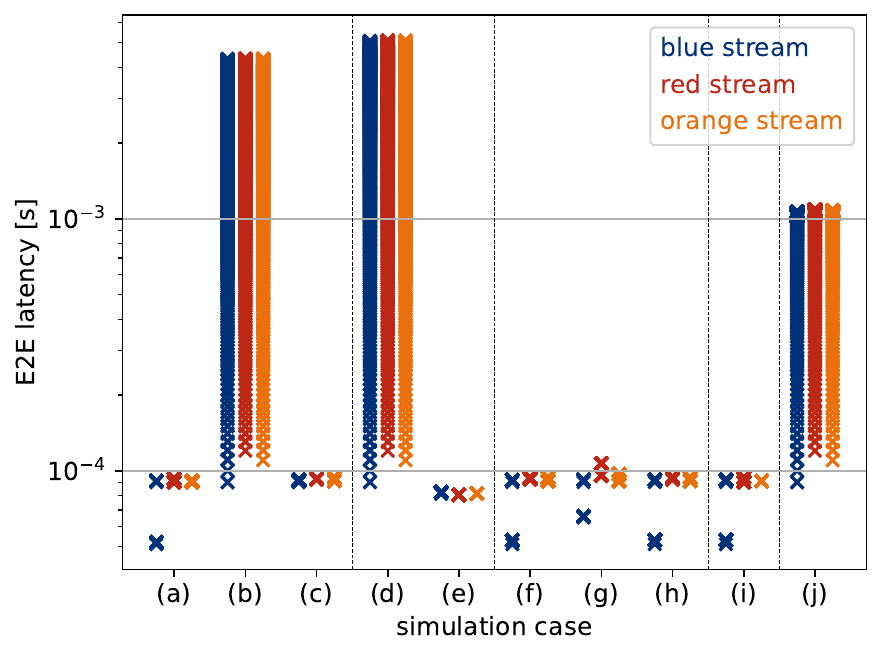}
    \caption{End-to-end latencies of the three streams of the adversarial traffic generation and proposed \ac{ATS} configurations in network A (and network C in case (e)).}
    \label{fig:small_latencies}
\end{figure}

Figure \ref{fig:small_latencies} presents the \ac{E2E} latencies of the blue, red and orange streams for the baseline cases in network A. 
For scaling, we limit the depiction to the first 1000 latency values, corresponding to the first \SI{7}{\ms} of the simulation time.
Values that are not depicted continue the trend of the values shown.

Baseline case \enum{a} simulates the networks without \ac{ATS} on any switch. 
The \ac{E2E} latencies are bounded for all streams.
All frames of the red and orange streams always have the same latency.
Frames of the blue stream have two distinct latencies:
In network A it depends on whether the frame takes the long or short path, in networks B and C it depends on whether the frame is delayed by the cross traffic.

\ac{ATS} is used on switch 2 in baseline case \enum{b}.
The \ac{E2E} latencies of all three streams increase over time.

Baseline case \enum{c} uses \ac{ATS} on switch 2, but the \ac{ATS} schedulers for the three streams are in different scheduler groups.
This is the solution proposed in \cite{tl-ncsjr-24}.
The \ac{E2E} latencies for all three streams are bounded. 
In contrast to case \enum{a}, there is only one value of the latency for the blue stream, because the frames with a shorter delay (short path or not delayed by cross-traffic), are then delayed by the shaping.
While this solution prevents unbounded latencies, its adoption requires a change of the standard. 
Furthermore, scheduler groups limit the number of FIFO queues needed to implement \ac{ATS}~\cite{ss-ubsjr-16}.

\subsection{Evaluation of \ac{ATS} Configurations}
Next, the four proposed solutions from Section \ref{sec:solutions} are implemented in the synthetic networks.
Figure \ref{fig:small_latencies} presents the first 1000 \ac{E2E} latencies for the blue, red, and orange stream in the following cases: 
\enum{d}- \enum{e} solution \ref{sec:solution_1} in networks A and C, and
\enum{f}- \enum{h} solution \ref{sec:solution_2}, \enum{i} solution \ref{sec:solution_3}, and \enum{j} solution \ref{sec:solution_4} in network A.

\subsubsection{Use \ac{ATS} on all hops}
\ac{ATS} is used on all switches in the networks in cases \enum{d} and \enum{e}.
The \ac{ATS} parameters are the same for all switches and streams:  $\ac{cir} = 20\,\unit{\mega\bit/\second}$, $\ac{cbs}=125\,\unit{\byte}$, and $\ac{mrt}$ is set to infinity. 
Network A, where \ac{FRER} is used, is presented in case \enum{d}.
The latencies of the three streams are unbounded, because the critical order of frames is caused by the merging.
In networks B and C (case \enum{e}), on the other hand, the latencies are bounded for all three streams, because the shaping on switch 1 puts the frames back to their original order of production.
Contrasting results in network A against those of network B and C shows that this solution is not always able to prevent unbounded latencies.

\subsubsection{Increase \ac{cir} or \ac{cbs}}
Cases \enum{f} to \enum{h} use \ac{ATS} only on switch 2.
In case \enum{f} both \ac{cir} and \ac{cbs} are doubled, case \enum{g} doubles only \ac{cir}, and case \enum{h} doubles only \ac{cbs}.

Case \enum{f}, which doubles both \ac{cir} and \ac{cbs}, results in \ac{E2E} latencies similar to case \enum{a}, where no \ac{ATS} is used.
The reason is that the parameters are set such that the shaping does not delay any frame.
In case \enum{g} only the value of \ac{cir} is doubled.
The minimum \ac{E2E} latency of the blue stream increases in comparison to case \enum{f}, because the shaping adds a small delay to the frames that take the short path, or a not delayed by cross traffic, respectively.
Frames of the red stream that need to wait for the delayed frame of the blue stream are also delayed.
Case \enum{h} doubles only the value of \ac{cbs}.
This configuration does not delay any frames due to the shaping.
Increasing the parameter values increases the bandwidth and burst size of the affected stream(s) on the subsequent hops, which needs to be considered in the network configurations. 

\subsubsection{Do not place \ac{ATS} behind \ac{FRER} merging}
\ac{ATS} is used on all switches except switch 2 in case \enum{i}.
The \ac{ATS} parameters are the same as for solution \ref{sec:solution_1}.
This case is only relevant in network A, because networks B and C do not use \ac{FRER}.
The results in Figure \ref{fig:small_latencies} show that the latencies of all three streams are bounded for this case, because there is no \ac{ATS} after the frames are put into the critical order.
This solution is only applicable when the merging points in the network are placed such that they separate the network into distinct before- and after-merging parts.
An alternative traffic shaping mechanism might be implemented after merging. 

\subsubsection{Utilize the \ac{mrt} Parameter}
The \ac{mrt} is set for all three \ac{ATS} schedulers on switch 2 in case \enum{j}.
When \ac{mrt} is set, there is an upper bound to the \ac{E2E} latencies of the streams, because the delay added by the shaping is limited.
Figure \ref{fig:small_latencies} shows the results with $\ac{mrt}=\SI{1}{\ms}$, the presented simulation time is \SI{14}{\ms} due to frame loss.
The \ac{E2E} latencies for all three streams increase until they reach the value of \ac{mrt}+\textit{network delay} and then stay the same.

\begin{figure}
    \centering
    \includegraphics[width=\linewidth, trim=0.1cm 0.3cm 0.1cm 0.2cm, clip=true]{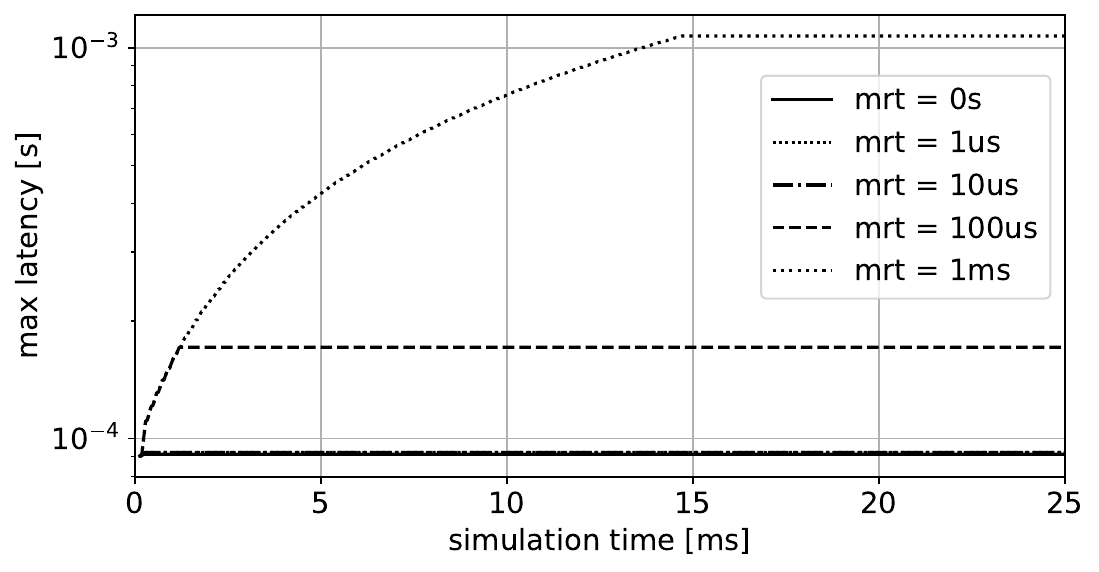}
    \caption{End-to-end latencies of the blue stream in network A with different values for \ac{mrt}. }
    \label{fig:maxlatencies_mrt}
\end{figure}

Figure \ref{fig:maxlatencies_mrt} illustrates this increase of \ac{E2E} latencies in detail.
Initially, the assigned eligibility times on switch 2 are within the range set by \ac{mrt}; the \ac{E2E} latencies increase during this phase.
When the assigned eligibility times exceed the range set by \ac{mrt}, frames are dropped and \ac{E2E} latencies stay stable.

For low values of \ac{mrt}, in the presented results 0\,\unit{\s} or 1\,\unit{\us}, frames of all streams are dropped due to timing imprecisions.
Otherwise, only frames of the blue stream are dropped.

While setting \ac{mrt} does prevent unbounded latencies, it enforces this by dropping frames, decreasing the reliability.

\section{Case Study: In-Vehicle Network} \label{sec:car}

We simulate a realistic \ac{IVN} with redundancy to show how the addition of \ac{ATS} after the stream recovery function leads to unbounded latencies.
We previously published this network as open source with a detailed explanation in~\cite{mhlks-fsaad-24}, and 
we make our \ac{ATS} configurations publicly available\footnote{\url{https://github.com/CoRE-RG/NIDSDatasetCreation}}.

The original network does not use \ac{ATS} but relies on \ac{CBS} to shape the streams.
Already when we replace \ac{CBS} with \ac{ATS} on the switches without adjusting any other configurations, the problem of unbounded \ac{E2E} latencies occurs. 
We apply our combined solutions \ref{sec:solution_1} \& \ref{sec:solution_3}, and solution \ref{sec:solution_2} to regain the bounded latencies of the original network.

\subsection{Baseline Network Setup}
\begin{figure}
    \centering
    \includegraphics[width=1\linewidth, trim=0.9cm 1.9cm 0.8cm 1.9cm, clip=true]{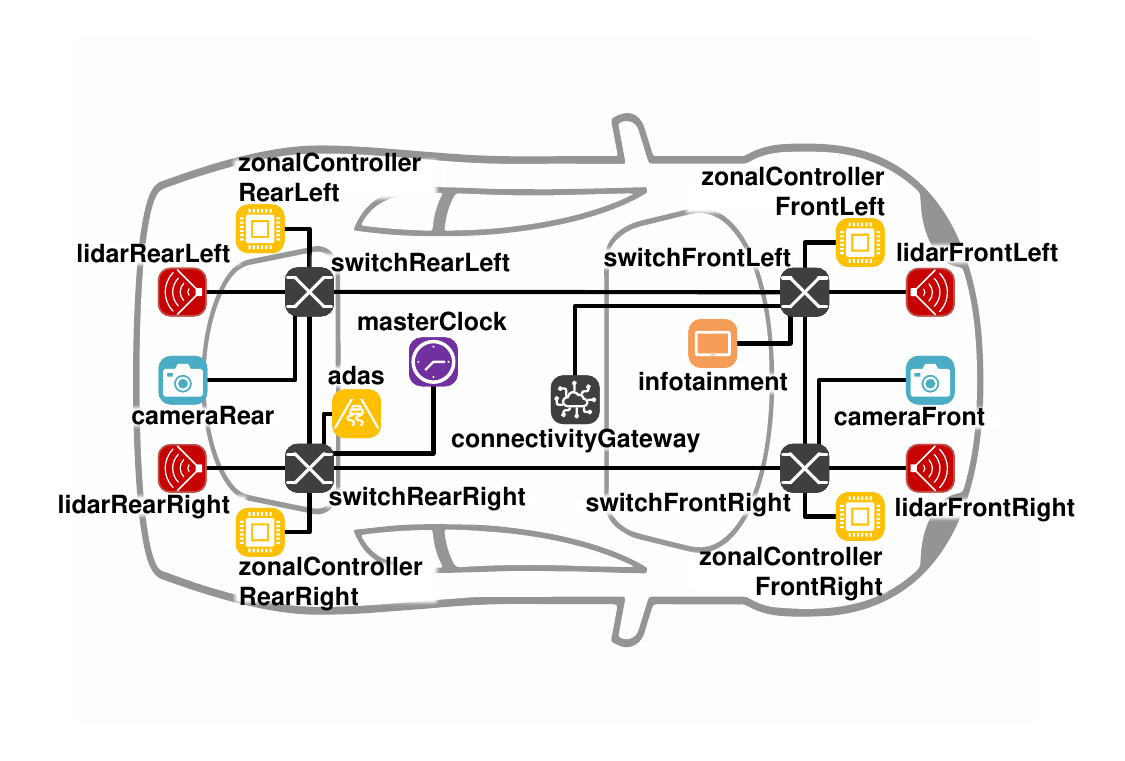}
    \caption{Future IVN canditate using a zonal topology with an Ethernet ring backbone and FRER for redundancy.}
    \label{fig:car_network}
\end{figure}

The realistic \ac{IVN} shown in Figure \ref{fig:car_network} has a zonal topology that employs a redundant ring-backbone with four switches.
Redundant streams are split on the first switch they enter and traverse the backbone both clockwise and counterclockwise, the stream recovery function is on the last switch they traverse.

Two video and four LIDAR streams are shaped with \ac{ATS} on their path to the \ac{ADAS} located in the rear right zone.
Video sources in the front right and rear left zone produce frames with a jitter of \SI{70}{\micro\second} and a bandwidth of \SI{176}{\mega\bit\per\second}.
LIDAR streams originate in all four zones with strict intervals and a bandwidth of \SI{104}{\mega\bit\per\second}.

All links have a bandwidth of \SI{1}{\giga\bit\per\second} with cross traffic on higher and lower priorities.
Furthermore, a \ac{TDMA} scheme on highest priority impacts the timing of all other priorities.
Unbounded latencies due to the combination of \ac{FRER} and \ac{ATS} can occur with at least three shaped streams~\cite{tl-ncsjr-24}.
Our \ac{IVN} has six shaped streams with the same destination, different path lengths, and heterogeneous traffic patterns due to jitter and cross traffic.
Rather than modifying these streams to create problematic traffic patterns, we rely on chance to produce them. 

We configure \ac{ATS} for the six streams on every switch and their source devices.
The \ac{cbs} is set to their respective frame sizes.
The \ac{cir} for the video streams is set to \SI{200}{\mega\bit\per\second} to meet \ac{E2E} latency requirements.
For LIDAR streams we set the \ac{cir} to their bandwidths. 
The \ac{ATS} schedulers on the switches have a \ac{mrt} set to \SI{50}{\micro\second}, it is large enough that there are no frame drops due to the \ac{ATS} scheduling.
We found these values in a preliminary parameter study where the values for \ac{cir} and \ac{mrt} were increased until the latency requirements were met.

\subsection{Evaluation of \ac{ATS} Configurations}
Our simulation case study covers three cases: 
\one A baseline that replaces \ac{CBS} with \ac{ATS} after the \ac{FRER} merger,
\two applies solutions \ref{sec:solution_1} and \ref{sec:solution_3} for the network configuration, and 
\three applies solution \ref{sec:solution_2} on \ac{ATS} schedulers after the merger

We simulate each case for \SI{10}{\second} simulation time.
Table \ref{tab:car_latencies} summarizes the minimum and maximum \ac{E2E} latencies of the six relevant streams. 
The minima do not change between the cases.
The differences between streams of the same type are due to their path lengths.
LIDAR 4 has the shortest path, with only one switch, and LIDAR 1 has the longest minimum path with three switches regardless of the direction.

\begin{table}
    \centering
    \caption{Min and max end-to-end latencies of video and LIDAR streams in the \ac{IVN} for different \ac{ATS} configurations.}
    \label{tab:car_latencies}
    \setlength{\tabcolsep}{5pt}
    \begin{tabularx}{\linewidth}{l rr p{1pt}rrp{1pt} rr}
        \toprule   
        Stream & \multicolumn{2}{c}{Baseline} & \multicolumn{4}{c}{Solution \ref{sec:solution_1} \& \ref{sec:solution_3}} & \multicolumn{2}{c}{Solution \ref{sec:solution_2}}   \\ 
        \cmidrule(rl){2-3} \cmidrule(rl){4-7} \cmidrule(rl){8-9} 
     & min            & max             & & min             & max              & & min             & max             \\ 
    \midrule
    video 1 & \SI{34}{\micro\second} & \SI{8.36}{\milli\second} & & \SI{34}{\micro\second} &  \SI{227}{\micro\second} & & \SI{34}{\micro\second}  & \SI{213}{\micro\second} \\
    video 2 & \SI{34}{\micro\second} & \SI{8.4}{\milli\second}  & & \SI{34}{\micro\second} &  \SI{256}{\micro\second} & & \SI{34}{\micro\second}  & \SI{221}{\micro\second} \\ 
    % \midrule
    LIDAR 1 & \SI{76}{\micro\second} & \SI{8.3}{\milli\second}  & & \SI{76}{\micro\second} &  \SI{120}{\micro\second} & & \SI{71}{\micro\second}  & \SI{124}{\micro\second} \\
    LIDAR 2 & \SI{46}{\micro\second} & \SI{8.28}{\milli\second} & & \SI{44}{\micro\second} &  \SI{110}{\micro\second} & & \SI{44}{\micro\second}  & \SI{103}{\micro\second} \\
    LIDAR 3 & \SI{44}{\micro\second} & \SI{8.29}{\milli\second} & & \SI{44}{\micro\second} &  \SI{104}{\micro\second} & & \SI{44}{\micro\second}  & \SI{106}{\micro\second} \\
    LIDAR 4 & \SI{28}{\micro\second} & \SI{8.25}{\milli\second} & & \SI{28}{\micro\second} &  \SI{51}{\micro\second}  & & \SI{28}{\micro\second}  & \SI{69}{\micro\second} \\
    \bottomrule
    \end{tabularx}
\end{table}

For the baseline, we replace \ac{CBS} with \ac{ATS} schedulers after the stream recovery function.
The per stream values for \ac{cir} and \ac{cbs} are the same in all switches.
The \ac{mrt} is set to infinity on all schedulers to ensure that the maximum \ac{E2E} latency is not restricted by frame drops.
After one second, the maximum \ac{E2E} latencies have increased to \SI{8.4}{\milli\second} for the video streams and \SI{8.3}{\milli\second} for the LIDAR streams.
Longer simulations lead to higher, unbounded, maximum \ac{E2E} latencies.

Applying solution \ref{sec:solution_1} \& \ref{sec:solution_3} placing all \ac{ATS} schedulers before the stream recovery function reduces the \ac{E2E} latencies.
The maximum \ac{E2E} delay is \SI{256}{\micro\second} for the video streams and \SI{120}{\micro\second} for the LIDAR streams.

Alternatively, we apply solution \ref{sec:solution_2} by doubling the \ac{cbs} of the LIDAR streams. 
Experiments showed that the \ac{cir} of the video streams is large enough to not cause unbounded latencies, therefore, only the \ac{cbs} of the LIDAR streams is increased.
We achieve a maximum \ac{E2E} latency of \SI{221}{\micro\second} for video streams, and \SI{124}{\micro\second} for the LIDAR streams. 

With this, we demonstrate that our solutions help regaining bounded latencies in the \ac{IVN} case study.
In the cases that apply our solution, longer simulations do not increase the maximum latencies, indicating upper bounds.

%% file: 05_conclusion.tex
\section{Conclusion and Outlook} \label{sec:conclusion}
TSN traffic shaping mechanisms, such as \ac{ATS}, are promising solutions for ensuring deterministic latencies in \acp{IVN}. 
However, related work has shown that \ac{ATS} can introduce unbounded latencies in networks with redundancy~\cite{tml-wcdjr-22} or non-FIFO behavior~\cite{tl-ncsjr-24}. In this work, we proposed configuration guidelines for \ac{ATS} that avoid these unbounded delays.

Using \ac{ATS} in every switch in a network can prevent frames from entering \ac{ATS} schedulers in a critical order. 
However, when using \ac{FRER}, this is insufficient. 
The \ac{ATS} parameters on switches after the stream recovery function must be adjusted to prevent shaping from introducing additional latencies. 
Our results indicate that it may be advisable not to use \ac{ATS} after the stream recovery function. 
Furthermore, while the \ac{mrt} parameter can limit delays introduced by \ac{ATS}, using it to reduce delays only shifts the problem toward frame loss.

We evaluated our solutions in a realistic \ac{IVN} scenario, where replacing \ac{CBS} with \ac{ATS} led to unbounded latencies.
Again, by ensuring \ac{ATS} was not applied after the recovery function and increasing the \ac{cbs} parameter for \ac{ATS} schedulers after the recovery function, we successfully prevented these latencies.
This provides a workaround for the interaction of \ac{ATS} and \ac{FRER} in \acp{IVN}.

This work gives rise to three future research directions. 
First, our solutions may be explored in other domains, such as industrial networks.
Second, a formal validation of the methods will increase the confidence in their effectiveness for real-world deployments.
Finally, a performance comparison of our \ac{ATS} configurations for the \ac{IVN} case study with other traffic shaping mechanisms could optimize configurations for \ac{IVN} scenarios.

%% file: main.bbl
\begin{thebibliography}{10}
\providecommand{\url}[1]{#1}
\csname url@samestyle\endcsname
\providecommand{\newblock}{\relax}
\providecommand{\bibinfo}[2]{#2}
\providecommand{\BIBentrySTDinterwordspacing}{\spaceskip=0pt\relax}
\providecommand{\BIBentryALTinterwordstretchfactor}{4}
\providecommand{\BIBentryALTinterwordspacing}{\spaceskip=\fontdimen2\font plus
\BIBentryALTinterwordstretchfactor\fontdimen3\font minus \fontdimen4\font\relax}
\providecommand{\BIBforeignlanguage}[2]{{%
\expandafter\ifx\csname l@#1\endcsname\relax
\typeout{** WARNING: IEEEtran.bst: No hyphenation pattern has been}%
\typeout{** loaded for the language `#1'. Using the pattern for}%
\typeout{** the default language instead.}%
\else
\language=\csname l@#1\endcsname
\fi
#2}}
\providecommand{\BIBdecl}{\relax}
\BIBdecl

\bibitem{wtm-avnjr-21}
J.~Walrand \emph{et~al.}, ``{An Architecture for In-Vehicle Networks},'' \emph{IEEE Trans. Veh. Technol.}, vol.~70, pp. 6335--6342, Jul. 2021.

\bibitem{psjtx-svtjr-23}
Y.~Peng \emph{et~al.}, ``{A Survey on In-Vehicle Time-Sensitive Networking},'' \emph{IEEE Internet of Things Journal}, vol.~10, pp. 14\,375--14\,396, 2023.

\bibitem{hmks-snsti-19}
T.~H{\"a}ckel \emph{et~al.}, ``{Software-Defined Networks Supporting Time-Sensitive In-Vehicular Communication},'' in \emph{IEEE 89th VTC2019-Spring}, Apr. 2019, pp. 1--5.

\bibitem{mhlks-fsaad-24}
P.~Meyer \emph{et~al.}, ``{A Framework for the Systematic Assessment of Anomaly Detectors in Time-Sensitive Automotive Networks},'' in \emph{IEEE 15th VNC}, May 2024, pp. 57--64.

\bibitem{ieee8021q-22}
{IEEE}, ``{Standard for Local and Metropolitan Area Networks--Bridges and Bridged Networks},'' IEEE, Std 802.1Q-2022, Dec. 2022.

\bibitem{ntaws-pcijr-19}
A.~Nasrallah \emph{et~al.}, ``{Performance Comparison of {IEEE} 802.1 {TSN} Time Aware Shaper ({TAS}) and Asynchronous Traffic Shaper ({ATS})},'' \emph{IEEE Access}, vol.~7, pp. 44\,165--44\,181, 2019.

\bibitem{flgx-sacjr-20}
B.~Fang \emph{et~al.}, ``{Simulative Assessments of Credit-Based Shaping and Asynchronous Traffic Shaping in Time-Sensitive Networking},'' in \emph{12th ICAIT}, 2020, pp. 111--118.

\bibitem{zlbpy-sttjr-21}
Z.~Zhou \emph{et~al.}, ``{Simulating TSN Traffic Scheduling and Shaping For Future Automotive Ethernet},'' \emph{Journal of Communications and Networks}, vol.~23, pp. 53--62, 2021.

\bibitem{mvghg-vcbjr-23}
L.~Maile \emph{et~al.}, ``{On the Validity of Credit-Based Shaper Delay Guarantees in Decentralized Reservation Protocols},'' in \emph{ACM 31st RTNS}, 2023, pp. 108--118.

\bibitem{tml-wcdjr-22}
L.~Thomas \emph{et~al.}, ``{Worst-Case Delay Bounds in Time-Sensitive Networks With Packet Replication and Elimination},'' \emph{IEEE/ACM Trans. Netw.}, vol.~30, pp. 2701--2715, 2022.

\bibitem{tl-ncsjr-24}
L.~Thomas and J.-Y. Le~Boudec, ``Network-calculus service curves of the interleaved regulator,'' \emph{Performance Evaluation}, vol.~166, 2024.

\bibitem{hmks-stsdn-22}
T.~H{\"a}ckel \emph{et~al.}, ``{Secure Time-Sensitive Software-Defined Networking in Vehicles},'' \emph{IEEE Trans. Veh. Technol.}, pp. 1--16, 2022.

\bibitem{mhrks-nadct-21}
P.~Meyer \emph{et~al.}, ``{Network Anomaly Detection in Cars: A Case for Time-Sensitive Stream Filtering and Policing},'' \emph{Computer Networks}, vol. 255, Dec. 2024.

\bibitem{ieee8021dg-21}
{IEEE}, ``{Draft Standard for Local and Metropolitan Area Networks - Time-Sensitive Networking Profile for Automotive In-Vehicle Ethernet Communications},'' IEEE, Std. 802.1DG, Draft 1.4, Dec. 2021.

\bibitem{ss-ubsjr-16}
J.~Specht and S.~Samii, ``{Urgency-Based Scheduler for Time-Sensitive Switched Ethernet Networks},'' in \emph{28th ECRTS}, Jul. 2016, pp. 75--85.

\bibitem{ieee8021cb-17}
{IEEE}, ``{Standard for Local and metropolitan area networks -- Frame Replication and Elimination for Reliability},'' IEEE, Std. 802.1CB-2017, Jul. 2017.

\bibitem{yi-qeajr-24}
A.~Yoshimura and Y.~Ito, ``{QoS evaluation of ATS in IEEE 802.1 TSN on in-vehicle Ethernet by comparing with CBS and TAS},'' \emph{IEICE Communications Express}, 2024.

\bibitem{msmb-lbbjr-18}
E.~Mohammadpour \emph{et~al.}, ``{Latency and Backlog Bounds in Time-Sensitive Networking with Credit Based Shapers and Asynchronous Traffic Shaping},'' in \emph{IEEE 30th {ITC}}, Sep. 2018.

\bibitem{hlxf-dbajr-20}
H.~Hu \emph{et~al.}, ``{The Delay Bound Analysis Based on Network Calculus for Asynchronous Traffic Shaping under Parameter Inconsistency},'' in \emph{{IEEE} 20th {ICCT}}, Oct. 2020.

\bibitem{l-ttrjr-18}
J.-Y. Le~Boudec, ``{A Theory of Traffic Regulators for Deterministic Networks With Application to Interleaved Regulators},'' \emph{IEEE/ACM Trans. Netw.}, vol.~26, pp. 2721--2733, Dec. 2018.

\bibitem{omnetpp}
{OpenSim Ltd.}, ``{OMNeT++}.'' [Online]. Available: \url{https://omnetpp.org/}

\bibitem{inet-framework}
------, ``{INET Framework}.'' [Online]. Available: \url{https://inet.omnetpp.org/}

\end{thebibliography}
